\documentclass[sigconf,authorversion,nonacm,natbib=false]{acmart}

\usepackage{bm}
\usepackage{gensymb}
\usepackage{bbm}
\usepackage{booktabs}
\usepackage{pifont}
\usepackage{enumitem}
\usepackage{algorithmic}
\usepackage{graphicx}
\usepackage{hyperref}[url]
\usepackage{multirow}
\usepackage{textcomp}
\usepackage{makecell}
\usepackage{tabu}

\RequirePackage[
  datamodel=acmdatamodel,
  style=acmnumeric,
  ]{biblatex}

\definecolor{lightblue}{rgb}{0.9, 0.9, 1} 

\copyrightyear{2024}
\acmYear{2024}
\setcopyright{acmlicensed}\acmConference[ICMI '24]{INTERNATIONAL CONFERENCE ON MULTIMODAL INTERACTION}{November 4--8, 2024}{San Jose, Costa Rica}
\acmBooktitle{INTERNATIONAL CONFERENCE ON MULTIMODAL INTERACTION (ICMI '24), November 4--8, 2024, San Jose, Costa Rica}
\acmDOI{10.1145/3678957.3685722}
\acmISBN{979-8-4007-0462-8/24/11}

\begin{document}

\title{NapTune: Efficient Model Tuning for Mood Classification using Previous Night's Sleep Measures along with Wearable Time-series}

\author{Debaditya Shome, Nasim Montazeri Ghahjaverestan, Ali Etemad}
\affiliation{%
  \institution{Queen's University}
  \country{Canada}
}

\renewcommand{\shortauthors}{Shome et al.}

\renewcommand{\shorttitle}{NapTune}

\begin{abstract}
  Sleep is known to be a key factor in emotional regulation and overall mental health. In this study, we explore the integration of sleep measures from the previous night into wearable-based mood recognition. To this end, we propose NapTune, a novel prompt-tuning framework that utilizes sleep-related measures as additional inputs to a frozen pre-trained wearable time-series encoder by adding and training lightweight prompt parameters to each Transformer layer. Through rigorous empirical evaluation, we demonstrate that the inclusion of sleep data using NapTune not only improves mood recognition performance across different wearable time-series namely ECG, PPG, and EDA, but also makes it more sample-efficient. Our method demonstrates significant improvements over the best baselines and unimodal variants. Furthermore, we analyze the impact of adding sleep-related measures on recognizing different moods as well as the influence of individual sleep-related measures.
\end{abstract}

\keywords{Mood Recognition, Affective Computing, Sleep, Deep Learning}

\begin{teaserfigure}
\centering
  \includegraphics[width=0.85\textwidth]{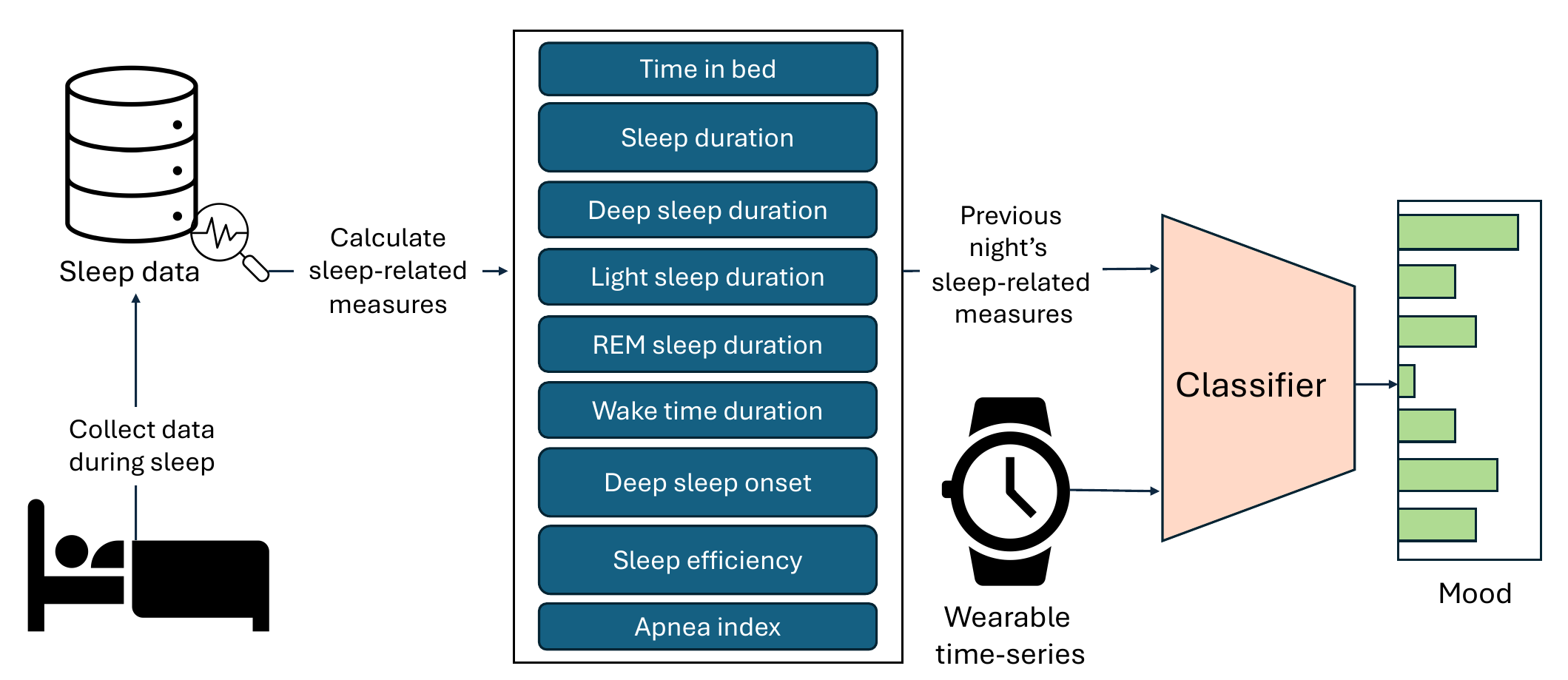}
  \caption{Overview of our proposed pipeline.}
  \Description{}
  \label{fig:teaser}
\vspace{10mm}
\end{teaserfigure}

\maketitle
\section{Introduction}
Affect recognition is a pivotal area in human-computer interaction, enabling a range of real-world applications \cite{calvo2010affect} ranging from healthcare to user experience. Specifically, in healthcare, it offers valuable insights into mental health treatments and stress management \cite{taylor2017personalized}. In education, it can assess students' engagement and stress levels, tailoring learning experiences accordingly \cite{kapoor2005multimodal}. In the workplace, affect recognition can enhance employee well-being by identifying stress or dissatisfaction \cite{mantello2023emotional}. And finally, for user experience, affective computing paves the way for more intuitive and responsive products, from smart homes that adjust environments based on the occupants' moods to entertainment systems that adapt to user reactions \cite{kaushik2022isecurehome}.

Wearable devices have been widely adopted for affect recognition due to their ability to continuously and unobtrusively monitor physiological signals \cite{schmidt2019wearable}. In particular, wearable signals such as electrocardiography (ECG) \cite{hasnul2021electrocardiogram, ross2023unsupervised}, photoplethysmography (PPG) \cite{lee2019fast}, and Electrodermal Activity (EDA) \cite{shukla2019feature} have been the focus of numerous studies within this field. ECG measures the heart's electrical activity and can therefore capture changes in the emotional state of users. PPG detects blood volume changes throughout the body, offering insight into the physiological responses to emotions. 
EDA measures the electrical conductance of the skin, which dynamically fluctuates with sweat gland activity which is influenced by sympathetic nervous system activations. With advances in machine learning, these signals have been used to develop models capable of accurately recognizing affective states in real-time \cite{jiao2020real}.

Mood and emotion are two common affect categories that are sometimes used interchangeably. In contrast to emotion, mood is an affective category that lasts for a long time, even up to several hours or days, and is thus harder to recognize from real-time data \cite{tizzano2020deep}. While real-time wearable signals provide valuable information about a subject's affect state, they sometimes do not encapsulate historical information or external factors that may have a strong impact on mood. In particular, sleep-related measures stand out as such a salient external factor known to influence mood \cite{sano2015prediction, ten2022sleep}. These measures may include `time in bed', `sleep duration', `deep sleep duration', `light sleep duration', `REM sleep duration', `wake time', `deep sleep onset', `sleep efficiency', and `Apnea index'.
As a critical component of daily life, sleep impacts emotional regulation and mental health, thereby shaping subsequent affective states. The relationship between sleep patterns and mood dynamics is well-studied in the literature \cite{wong2013interplay, kalmbach2014interplay, konjarski2018reciprocal}, with sleep disturbances known to aggravate stress and irritability \cite{whiting2023associations}. Conversely, a pattern of restorative sleep is often correlated with enhanced mood stability and cognitive functioning \cite{leong2023understanding}.

A thorough investigation of the literature in this area demonstrates that despite the relationship between sleep measures and mood being well-established in prior works, previous night's sleep measures have not been incorporated into automated mood recognition models as a complementary modality to wearable signals. Moreover, what makes this issue even more challenging is the scarcity of public datasets that contain the previous night's sleep measures \textit{paired} with wearable time-series from the \textit{same users} over a period of time. This in turn inhibits the use of standard training frameworks from being effective in enabling a multimodal setup for both sleep and wearable time-series simultaneously. 

In this paper, to address the challenges above, we propose NapTune, a framework that integrates previous night's sleep measures as a complementary source of information into mood recognition from various wearable time-series (ECG, PPG, and EDA). First, our model uses an unimodal Transformer-based encoder to extract effective representations from the wearable time-series. Next, to enable our model to incorporate and analyze the previous night's sleep-related measures as an additional input, we freeze the Transformer encoder and add a small number of learnable parameters into each layer, which we then train for mood recognition. This enables our model to effectively utilize sleep-related measures alongside wearable time-series for mood recognition, with little requirement for paired training data. Extensive experiments demonstrate that incorporating previous night's sleep-related measures leads to an increase in mood recognition performance with up to 8\% in F1 score. In summary, we make the following contributions. 

\noindent \textbf{(1)} We propose NapTune, an efficient tuning framework for adapting a frozen wearable time-series encoder to utilize the previous night's sleep-related measures as an additional input modality. Our framework demonstrates strong performances for mood classification based on sleep-related measures and wearable time-series, outperforming various multimodal baselines.

\noindent \textbf{(2)} For training, our framework requires minimal data in the form of wearable time-series paired with previous night's sleep measures from the same users, overcoming the problem of insufficient paired training data.
    
\noindent \textbf{(3)} We study the effect of the previous night's sleep-related measures in aiding mood recognition from wearable time-series. Particularly, we train and evaluate models for mood recognition with ECG, PPG, and EDA,  with and without the use of sleep measures. Our findings show an increase of 9\% to 11\% in F1 when utilizing the sleep-related measures as an additional modality. 

\section{Related Work}
In this section, we review three key areas relevant to our work. First, we review affect recognition studies from wearable time-series. Next, we present a review of prior works focusing on the relationship between sleep and affect. Finally, we conclude this section with a thorough review of recent works on prompt-tuning, a technique that is central to our proposed framework.

\subsection{Wearable affect recognition} 

Several recent studies have demonstrated the potential of wearable time-series for affect recognition. In \cite{zenonos2016healthyoffice}, a framework for personalized and generalized mood recognition in a workplace setting was proposed using smartphone-based wearable sensors including ECG, PPG, 3-axis acceleration, and skin temperature. Several ML algorithms were evaluated to find a bagged ensemble of decision trees as the best performing solution. In \cite{zhu2015naturalistic}, a novel pipeline using popular wearables was developed to recognize daily activities and a regression model was trained for mood assessment. The method was tested in a real-world study with 18 users over 93 user-days and achieved mood inference with a mean absolute error of 0.24 $\pi$ radians on the Circumplex Model of Affect. 
In \cite{sarkar2020self}, a self-supervised learning method was introduced for emotion recognition using ECG time-series, that achieves comparable or superior results to fully-supervised methods on the SWELL \cite{koldijk2014swell} and AMIGOS \cite{miranda2018amigos} datasets. In \cite{exler2016wearable}, a study was conducted on combining smartphone and smartwatch data to assess mood. 
The results demonstrated the significance of temporal features and heart rate obtained from such wearable time-series. In \cite{zhu2023electrodermal}, an emotion recognition system was developed using spectrogram representations of EDA time-series with a combination of CNN and Bi-GRU for feature learning and classification. The model was evaluated on the AMIGOS \cite{miranda2018amigos} dataset demonstrating high classification accuracies of 83.4\% for arousal and 81.2\% for valence.

\subsection{Sleep and affect}
Although the aim of this paper is to study the impact of sleep as an additional modality to improve mood recognition, in this subsection, we review the prior works that study the overall link between sleep and affect, i.e., how sleep impacts affect and vice-versa. In \cite{ten2022sleep}, a conceptual review was conducted to explore the relationship between sleep and affect. A granular framework was proposed to deconstruct sleep and affect into three dimensions: domains, methods, and timescales. This framework was applied to systematically review empirical studies from PubMed, focusing on associations between various aspects of sleep and affect. The study found evidence of links between sleep disturbances, sleep duration, and affect, but noted that evidence was often inconclusive or sparse for other aspects. In \cite{konjarski2018reciprocal}, the reciprocal relationships between daily sleep and mood were examined, focusing on real-world studies. Electronic databases were utilized to collect studies that investigated daily associations between sleep and mood using ambulatory diary techniques. The findings of this study supported a reciprocal relationship between subjective sleep variables (quality, duration, latency) and daytime affective states, emphasizing the potential clinical importance of sleep disturbance in predicting and preventing psychopathology. 

In \cite{kalmbach2014interplay}, a 2-week study was conducted to explore the interplay between daily affect and sleep in young women. Using daily sampling, they investigated how variations in positive and negative affect influenced self-reported sleep-onset latency, sleep duration, and sleep quality. The study revealed significant associations between sleep and emotions, where sadness and serenity were found to be strong predictors of sleep-related measures, and better sleep quality predicted greater happiness the next day. In \cite{cellini2017perceived}, the interplay of neuroticism, affect, and hyperarousal with perceived sleep quality was explored. Using an online survey, the impact of these factors on sleep quality was assessed among 498 Italian participants. The study found that neuroticism was the primary personality predictor of poor sleep quality. Additionally, hyperarousal and positive affect also significantly predicted good sleep quality. In \cite{krizan2023emotions}, a systematic review and meta-analysis was conducted to evaluate the causal impact of emotions on sleep. 31 experimental studies focusing on the effects of emotion inductions on various sleep-related measures were analyzed. The study found a moderately significant effect of emotion inductions on delayed sleep onset latency, but no consistent impacts on other sleep-related measures.

\vspace{-1em}
\subsection{Prompt tuning}
Prompt tuning was first introduced as a method for adapting large pre-trained language models to specific tasks using soft prompts learned through backpropagation \cite{lester2021power}. This technique allows a large frozen model to be used across multiple tasks without significant computational overhead.
Inspired by the success of this approach in large language models, prompt tuning has since been further adopted for other domians including vision and time-series. In \cite{jia2022visual}, Visual Prompt Tuning (VPT) was introduced as an efficient method for adapting large-scale Transformer models in vision tasks. VPT incorporates a small amount of trainable parameters into the input space while keeping the model backbone frozen, leveraging the advantages of large pre-trained models without the need for extensive retraining. 
Similarly, prompt-tuning was recently adopted for several time-series tasks demonstrating promising results. In \cite{cao2023tempo}, a novel framework named TEMPO was proposed as a prompt-based generative pre-trained Transformer for time-series forecasting. TEMPO addresses the challenges in time-series forecasting by integrating decomposition of complex interactions between trend, seasonal, and residual components, and by introducing selection-based prompts to adapt to non-stationary time-series. The framework demonstrated superior performance over state-of-the-art methods in various benchmark datasets. In \cite{xue2023promptcast}, PromptCast, a novel approach to time-series forecasting utilizing a prompt-based paradigm, was introduced. It transforms numerical data into prompts, leveraging pre-trained language models for forecasting.

\begin{figure*}[htbp!]
    \centering
    \includegraphics[width=0.8\textwidth]{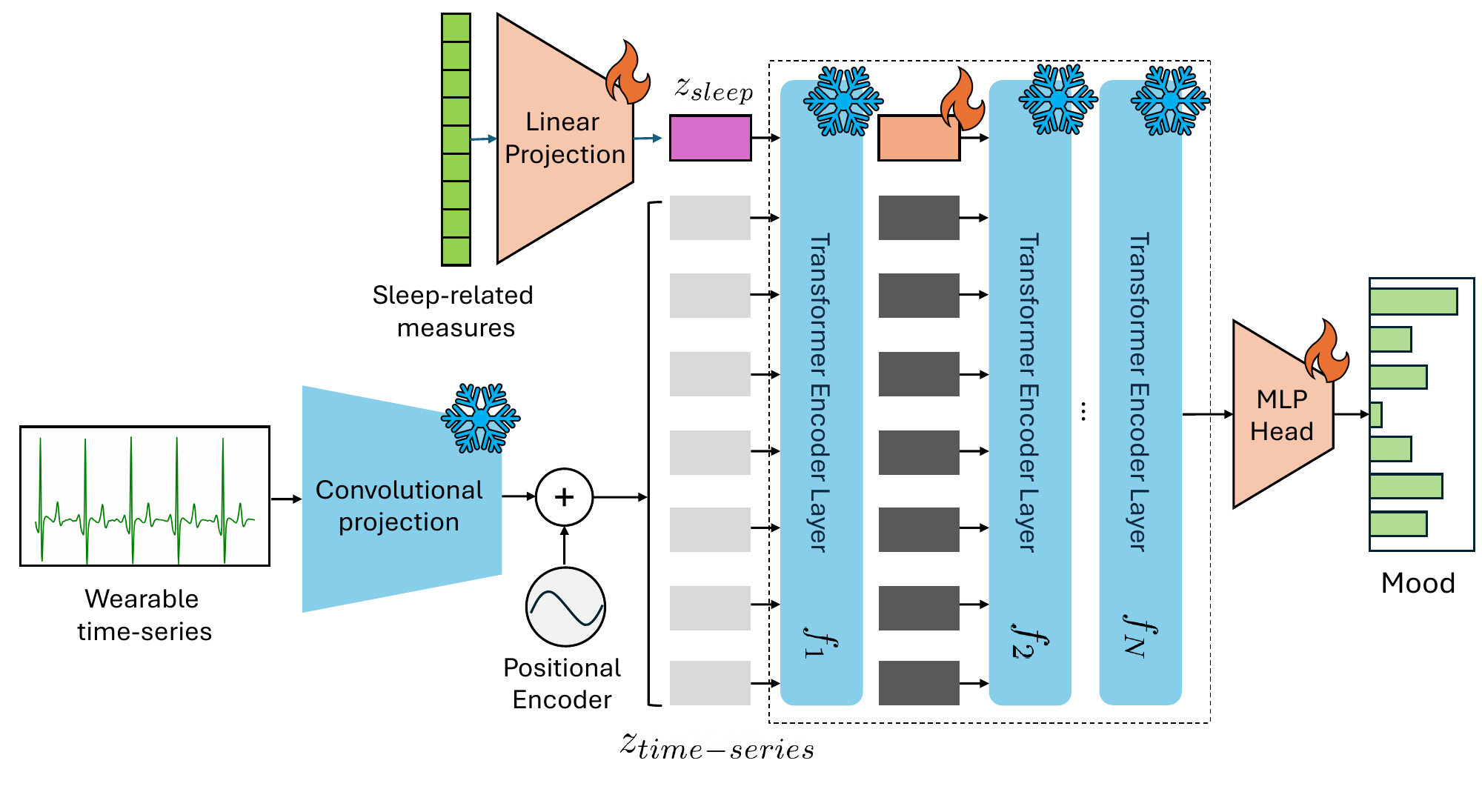}
    \caption{Our proposed NapTune framework.}
    \label{fig:framework}
\end{figure*}

In \cite{liu2021p}, PT-Tuning was introduced as an approach that enhances time-series forecasting by bridging the gap between masked reconstruction and forecasting through prompt token tuning. The unique aspect of PT-Tuning is its integration of a few trainable prompt tokens, while all other pre-trained parameters remain frozen. This effectively addresses the issue of task difficulty variation. The approach demonstrated remarkable performance improvements over other methods in extensive experiments with real-world datasets. In \cite{wang2023prompt}, a method called POND was introduced for multi-source time-series domain adaptation using prompt tuning. This method aims to tackle three challenges: the exploitation of domain-specific information for domain adaptation, the capturing of dynamic domain-specific information, and the assessment of learned domain-specific data. POND uses prompts to capture both common and specific information across domains, introduces a conditional module for generating dynamic prompts for each source domain, and employs specific criteria for effective prompt selection. In \cite{chang2023speechprompt}, SpeechPrompt v2 was introduced, which is a prompt tuning framework for various speech classification tasks. This approach uses generative spoken language models with trainable prompt vectors, allowing efficient adaptation to different languages and tasks with minimal parameter updates. The framework introduces a learnable verbalizer, enhancing its adaptability and performance across a wide range of speech classification tasks, including emotion recognition and language identification, while maintaining computational efficiency.

\section{Method}

\subsection{Problem and solution overview}
We aim to develop a framework to enable a pre-trained wearable time-series encoder to utilize the users' previous night's sleep-related measures as an auxiliary input to enhance mood recognition given a small training dataset of paired data. Inspired by recent advances in prompt tuning in large language foundation models, we propose an efficient tuning solution to address this problem. Our method includes two stages. First, we train a Transformer-based wearable time-series encoder using self-supervised pre-training over a large corpus of unlabeled unimodal wearable time-series. Next, we freeze the weights of this pre-trained encoder and add a linear projection to include sleep-related measures, and additional learnable parameters in each Transformer layer for efficient training without a high computational overhead. Figure \ref{fig:framework} depicts an overview of our approach.

\subsection{Proposed approach}

\subsubsection{Wearable time-series encoder}
\begin{figure*}[htbp!]
    \centering
    \includegraphics[width=0.75\textwidth]{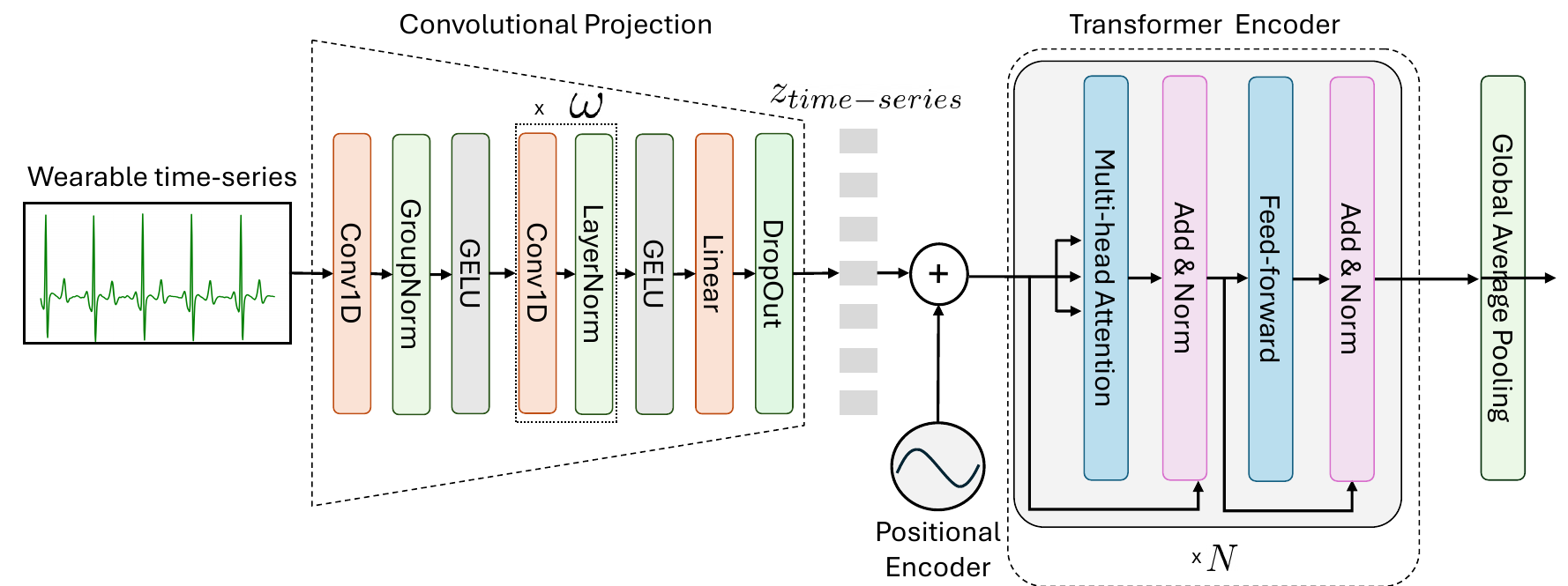}
    \caption{The architecture of the wearable time-series encoder.}
    \label{fig:architecture}
\end{figure*}

Inspired by \cite{behinaein2021transformer} we adopt a CNN-Transformer model as an encoder to learn the wearable time-series. 
The detailed architecture of this encoder is presented in Figure \ref{fig:architecture}. 
As shown in the figure, the encoder consists of a convolutional projection module with a series of $\omega$ convolution blocks. The initial convolutional layer uses a kernel size of $k_l$ (long kernel) and a stride of $s_l$ (long stride). This layer is specifically designed to better capture the global patterns in the input time-series. The output from this layer undergoes normalization using GroupNorm, followed by GELU activation. The subsequent convolutional layers follow a similar pattern where each of these layers use smaller kernels of size $k_s$ (short kernel) and a stride of $s_s$ (short stride), which focus on extracting more detailed local features from the input time-series. We apply LayerNorm to the extracted convolutional features and apply a GELU-activated linear layer to project the output of this convolutional projection module into a $L \times D$ dimensional time-series embedding $z_{time-series}$. 
Next, we represent $z_{time-series}$ as a set of $L$ tokens each with a dimension $D$, as
\begin{equation}
    z_{time-series} = \{z_{i} \in \mathbb{N} | 1 \leq i < L\}.
    \label{eqn:z_timeseries}
\end{equation} 

Next, in order to preserve sequential information we use a Positional Encoder. Given a time index $pos$ and dimension index $i$, the Positional Encoder generates a positional embedding $\Pi$ defined as
\begin{equation}
    \Pi(\text{pos}, 2i) = \sin\left(\frac{\text{pos}}{10000^{2i/D}}\right),
    \Pi(\text{pos}, 2i+1) = \cos\left(\frac{\text{pos}}{10000^{2i/D}}\right)
\end{equation}

Lastly, the model consists of a Transformer encoder, which we refer to as $f$, that consists of $N$ layers with an embedding dimension of $D$. Each layer consists of a multi-head attention module, LayerNorm, and GELU-activated feedforward layers, as shown in Figure \ref{fig:architecture}. Each multi-head attention module performs scaled-dot product attention in parallel over multiple attention heads by utilizing the layer's input representation as query, key, and value. Specifically, the scaled-dot product attention operation involves calculating the dot product between scaled query ($Q$) and key ($K$) matrices, which is multiplied with a pre-defined mask of attention weights. Then, the softmax operation is applied to derive the attention weights and subsequently multiplied with the value matrix ($V$) to obtain the final attention output represented as 
\begin{equation}
    \text{Attention}(Q, K, V) = \text{Softmax}\left(\frac{Q \cdot K^T}{\sqrt{d}} \ast \textrm{Mask}\right) V ,
\end{equation}
where, $d$ represents the dimension of $K$, $Q$, and $V$, which is used as a scaling factor to prevent vanishing gradients.

\subsubsection{Pre-training} To pre-train the wearable time-series encoder $f$ to learn strong representations from unlabeled time-series, we pre-train this module using SimCLR \cite{chen2020simple}. SimCLR is a widely used contrastive self-supervised learning framework \cite{chen2020simple, soltanieh2023distribution}. Given each input time-series $x$, we create two views $x_1$ and $x_2$ using two distinct stochastic augmentations $t_1$ and $t_2$. First, the encoder backbone extracts embedding $h$ from the augmented views of each time-series. Next, a projector network $g$ with a single linear layer projects the embedding into $z = g(h)$. In a minibatch of $M$ samples resulting in $2M$ pairs, we obtain one positive pair $(p, q)$, and the rest $2(M-1)$ as negative pairs. We use the contrastive loss function \cite{sohn2020fixmatch}, which can be defined as
\begin{equation}
    \ell_{p,q} = -\log \frac{\exp(\mathrm{sim}(\bm z_p, \bm z_q)/\tau)}{\sum_{k=1}^{2M} 1_{k \neq i}\exp(\mathrm{sim}(\bm z_p, \bm z_k)/\tau)},
\end{equation}
where $\mathrm{sim}$ indicates pairwise dot product similarity and $\tau$ is the temperature parameter. Both $t_1$ and $t_2$ are sampled from the same family of augmentations which are suitable for temporal time-series, as described below. 

\noindent \textit{Time Warping.} In this augmentation, we split the time-series $x$ into $r$ segments $x = [x_{1}, ..., x_{r}]$. We randomly select $\frac{r}{2}$ segments and apply 1D interpolation-based time-warping to stretch them by a factor of $\sigma$\%, while squeezing the remaining segments by the same factor. Finally, we concatenate the segments and apply zero padding if the resulting length is an odd number.
    
\noindent \textit{Gaussian Noise.} In this augmentation, gaussian noise is added to the time-series $x$. The noise standard deviation is randomly selected within a predetermined range, defined by the product of the time-series's standard deviation and specified minimum and maximum signal-to-noise ratios (SNR). Gaussian noise is then generated according to this noise standard deviation and added to $x$, resulting in a controlled level of random noise added to the time-series.
    
\noindent \textit{Random Scaling.}  This augmentation involves altering the amplitude of the time-series $x$. It is achieved by multiplying $x$ with a scaling factor $\alpha$, where $\alpha$ is a positive value randomly chosen from a predefined range.

\subsection{Tuning}
\label{ref:prompt-tuning}
Once the wearable time-series encoder is pre-trained, we freeze the weights of the convolutional projection module and the transformer encoder $f$. We add a linear projection module to encode sleep-related measures into a $D$-dimensional representation $z_{sleep}$. 
Using the wearable time-series encoder, we obtain $z_{time-series}$ using Equation \ref{eqn:z_timeseries}. Next, we concatenate $z_{sleep}$ with $z_{time-series}$ which results in $z_0$, a $D$-dimensional set of $(L+1)$ as input tokens for $f$.
Lastly, we add learnable parameters into each transformer layer of $f$ which are updated using backpropagation. Here, $f$ has $N$ Transformer encoder layers which can be represented as 
\begin{equation}
z_{n+1} = 
\begin{cases}
f_1 (z_1), & \text{if}\ n = 1 \\
f_{n} (\mathrm{concat}[P_n, z_n]), & \text{otherwise},
\end{cases}
\end{equation}
where $P_{n}$ refers to a set of unique learnable parameters at layer $n$.
We train the model using the Binary Cross Entropy (BCE) loss function for our downstream task of multi-label mood classification. Specifically, BCE loss can be defined as 
\vspace{-0.5em}
\begin{equation}
    \mathrm{BCE}(p,\tilde{p}) = -\frac{1}{C}\sum_{j=1}^{C}[p_j\log{\tilde{p}_j} + (1-p_j)\log{(1-\tilde{p}_j)}],
\end{equation}
where, $C$ is the number of classes, and $p$ and $\tilde{p}$ represent the actual mood labels and the predicted logits by our model, respectively.

\section{Experiment Setup}
\subsection{Datasets}

Here, we describe the details of the datasets used in this paper. First, we use a collection of datasets for pre-training the wearable time-series, as well as a downstream mood classification dataset which we use for evaluating our proposed method.

\subsubsection{Pre-training datasets} The following datasets are used for pre-training the wearable time-series encoders.

    
\noindent \textbf{WESAD \cite{schmidt2018introducing}.} This dataset contains 24 hours of Lead-II ECG and PPG data, sampled at 700 Hz and 64 Hz respectively, from 15 participants. Annotations in the dataset include stress and various affect states, providing insights into emotional and stress-related physiological responses. We use this dataset to pre-train encoders for ECG, PPG, and EDA.

\noindent \textbf{MIMIC AFib \cite{bashar2019noise}.} This datasets is part of the larger MIMIC-III waveform collection \cite{johnson2016mimic} dataset. This subset includes data from 35 critically ill adults, including 19 with Atrial Fibrillation (AFib). Both ECG and PPG time-series are present, sampled at a frequency of 125 Hz.

We use this dataset to pre-train ECG and PPG encoders.

\noindent \textbf{CAPNO \cite{karlen2021capnobase}.} This dataset features approximately 5.6 hours of Lead-II ECG and PPG data, recorded at a 300 Hz sampling rate, from 42 subjects under clinical supervision. It extends the range of physiological data to include medically monitored scenarios. We use this dataset to pre-train both ECG and PPG models. 

\noindent \textbf{BIDMC \cite{pimentel2016toward}.} Collected from 53 ICU patients, the dataset contains nearly 7 hours of ECG (across multiple leads including Lead II, V, and AVR) and PPG recordings, each sampled at 125 Hz. We use this dataset to pre-train encoders for ECG and PPG.

\noindent \textbf{DALIA \cite{reiss2019deep}.} Comprising approximately 35 hours of data, this dataset includes PPG and Lead-II ECG recordings from 15 subjects performing everyday activities. The sampling rates for ECG and PPG are 700 Hz and 64 Hz, respectively.

In our study, we use this dataset to pre-train ECG and PPG encoders.

\noindent \textbf{\citeauthor{fekri2023regulation} \cite{fekri2023regulation}.} This dataset includes EDA signals collected using an Empatica E4 device during cognitive tasks involving auditory, gustatory, and olfactory stimulation. It is designed to explore the modulation of cognitive states and its effects on physiological responses. We utilize this to pre-train the EDA encoder.

\subsubsection{Downstream Dataset}
    
For our experiments on the downstream task of mood classification, we require a datasets that contains both wearable time-series data with mood labels, paired with previous night's sleep-related measures from the same subjects. To our knowledge, \textbf{ECSMP dataset \cite{gao2021ecsmp}} is the only dataset that contains such information. It comprises multiple physiological time-series, including ECG, PPG, and EDA collected from 89 healthy college students. The sampling rates of ECG, PPG, and EDA are 512 Hz, 64 Hz, and 4 Hz respectively. These signals have been collected during various states, such as resting, emotional induction and recovery, and cognitive assessments. Additionally, the dataset consists of multiple self-reported questionnaires, among which we utilize the Profile of Mood States (POMS) as annotations for mood classification. It includes seven mood labels represented by scores to categorize tension, anger, fatigue, depression, vigor, confusion, and esteem. We convert the scores to binary labels and formulate this task as a $7$-way multi-label classification problem. 

The dataset also contains sleep-related measures estimated from the ECG collected during the previous night's sleep from each subject. The authors used the cardiopulmonary coupling analysis algorithm \cite{thomas2005electrocardiogram} to estimate the sleep-related measures from the ECG. These sleep-related measures include:
\begin{itemize}[leftmargin=*]
    \item \textit{Time in bed:} This measures the total number of hours that the subject spent in bed. 
    \item \textit{Sleep duration:} This refers to the total number of hours the subject has slept.
    \item \textit{Deep sleep duration:} This refers to the total number of hours that the subject has been in deep sleep, the most restorative stage of sleep.
    
    \item \textit{Light sleep duration:} This refers to the total number of hours that the subject has been in light sleep, the transition period between wakefulness and deep sleep. 
    \item \textit{REM sleep duration:} This measures the total number of hours in Rapid Eye Movement (REM) sleep, the sleep stage characterized by rapid eye movements, vivid dreaming, and high brain activity, resembling wakefulness.
    \item \textit{Wake time:} This measures the total number of hours the subject has spent in bed without sleeping.
    \item \textit{Deep sleep onset:} This indicates the number of hours needed by the subject to transition from light sleep to deep sleep.
    \item \textit{Sleep efficiency:} This refers to the ratio of sleep duration to the total time spent in bed.
    \item \textit{Apnea index:} This refers to the average number of apnea occurrences per hour of sleep.
    
    \end{itemize}

\subsection{Data pre-processing}
We apply standard pre-processing steps \cite{pan1985real, Makowski2021neurokit} for ECG, PPG, and EDA time-series. The ECG signals are processed using a high-pass Butterworth filter with a 0.5 Hz cut-off frequency. For the PPG time-series, we apply a band-pass Butterworth filter with frequencies ranging from 0.5 to 8 Hz. We skip the filtering of EDA time-series due to the low sampling frequency of 4 Hz. Z-score normalization is applied on each type of time-series to adjust for individual subject-specific differences. After normalization, the time-series are scaled to fit within a [-1, 1] range using min-max scaling. The final step in our pre-processing is segmenting the ECG, PPG, and EDA time-series into 10-second windows.

\subsection{Baselines}
Here we describe the models that we use as baselines for comparison to our proposed method. To our knowledge, no prior work has performed mood classification with the aid of the previous night's sleep measures on this dataset. As a result, we create several baselines based on state-of-the-art wearable time-series encoders and pre-train each encoder in the same manner that our own model's encoders are pre-trained. To create these baselines, we use an encoder/projection model for the sleep measures and perform late-stage fusion with the representations of the wearable time-series encoders. 
For the sleep encoder models, we use a GELU-activated 2-layer linear projection module for the Transformer-based baseline, while we use a ReLU-activated 2-layer linear projection module for others. Finally, for late-stage fusion, we combine the sleep embeddings with the global-average pooled output embeddings from the wearable time-series encoder and pass the outcome to a sigmoid-activated linear layer to predict mood logits.
To ensure that the best possible performances are obtained from these baselines, we optimize the hyper-parameters to the best of our ability. Following we describe the architectures and hyper-parameters of these encoders.

\noindent \textbf{\citeauthor{chen2021signal} \cite{chen2021signal}.} This model employs a VGG-19 \cite{simonyan2014very} backbone which takes 2D Short-Time Fourier Transform (STFT) representations of the wearable time-series as inputs. The network comprises a series of convolutional layers, initially with 64 filters, doubling at each stage following a max-pooling layer, up to layers with 512 filters. Each convolutional block consists of layers in configurations of two or four, paired with max pooling. Convolutional layers use 3x3 kernels with batch normalization and ReLU activation.

\noindent \textbf{\citeauthor{zihlmann2017convolutional}\cite{zihlmann2017convolutional}} This model follows a Convolutional Recurrent neural network (CRNN) architecture that takes wearable time-series converted into logarithmic spectrograms as input. It comprises 3 convolutional blocks, with each applying a set of 5x5 convolutional filters followed by batch normalization and ReLU activation. The channels in these layers increase with each block. A 3-layer bidirectional LSTM processes the outputs from the convolutional layers, providing an aggregated feature vector.

\noindent \textbf{\citeauthor{hong2020holmes} \cite{hong2020holmes}} We adopt ResNet1D \cite{hong2020holmes} as our baseline, an adaptation of the ResNet model \cite{he2016deep} for 1D time-series. The network is comprised of 34 blocks. Each block consists of convolution layers with a kernel size of 64 and a stride of 2, followed by batch normalization ReLU activation, and a dropout of $20\%$. We use 64 as the number of base filters, which doubles at specified intervals across the network.

\noindent \textbf{\citeauthor{behinaein2021transformer} \cite{behinaein2021transformer}} We use the same pre-trained wearable signal encoder based on Transformer-based architecture used for our proposed method as a baseline.

\subsection{Evaluation Protocol}
We use a cross-subject 3-fold cross-validation scheme for evaluation. This involves dividing all the subjects into three distinct groups (folds), with each fold being iteratively used as the test set while the remaining two serve as the training set. Due to the inherent imbalance in the mood states represented in the dataset, we report both weighted F1 score and accuracy. 

To obtain the final results, we generally perform linear evaluation unless otherwise specified. For linear evaluation, we use transfer learning, where we first freeze the weights of the pre-trained wearable time-series encoders. Then, we train a linear classifier over the learned representations for mood recognition.
 
\subsection{Implementation Details}
We use 4$\times$NVIDIA A100 GPUs for self-supervised pre-training using a batch size of 2048, with AdamW optimizer and a learning rate of 1$\times 10^{-4}$ for 150 epochs.  For downstream experiments, we use 2$\times$NVIDIA A100 GPUs with a batch size of 1024. We use AdamW optimizer with $\mathrm{CosineWarmup}$, starting from a base learning rate of 1$\times 10^{-5}$ for 30 epochs. To ensure reproducibility we present all the hyper-parameters used in our work, in Table \ref{table:model_details}.

\begin{table}[t]
\centering
\caption{The details of our wearable time-series encoder.}
\label{table:model_details}
\small{
\begin{tabular}{l|l|l}
 \hline
 \textbf{Module} & \textbf{Parameter} & \textbf{Value} \\ 
 \hline
 Convolutional Projection & Activation & GELU \\
 & $k_{l} $ & 10\\
 & $s_{l} $ & 5\\
 & $k_{s} $ & 3\\
 & $s_{s} $ & 2\\
 & $\omega$ & 6\\
 & Linear dim. & 512 \\
 & DropOut & 0.2\\
 \hline
 Linear Projection & Activation & GELU\\
 & Linear 1 dim. & 128\\
 & Linear 2 dim. & 512\\
 \hline
 Transformer Encoder & Activation & GELU \\
 & $N$ & 6\\
 & $D$ & 512 \\
 & heads & 4 \\
 \hline
 Classification Block  & Activation & GELU \\
 & Number of layers & 2 \\
 & Dropout rate & 0.5\\

 \hline
\end{tabular}
}
\end{table}

\section{Results and Discussion}
First, we compare our proposed method against several baselines using linear evaluation, and present the results in Table \ref{table:multimodal}. We observe that our approach achieves the best performance for all three types of wearable time-series for both metrics (F1 and accuracy) by considerable margins. More specifically, for EDA, our method achieves the best results by  2\% improvement in F1 and 3\% improvement in accuracy over the best baseline \cite{behinaein2021transformer}. For PPG, our method leads to a significant increase of 12\% and 10\% in F1 and accuracy respectively over the best baseline. With ECG, we achieve a similar improvement of 11\% and 13\% in F1 and accuracy, respectively. Out of the three modalities, NapTune performs best with ECG achieving an F1 of 0.76 and an accuracy of 0.71. We believe that the relatively lower performance when using EDA is due to the low sampling rate and the presence of noise.

\begin{figure*}[t!]
    \centering

    \begin{tabular}{ccc}
        \includegraphics[width=0.3\linewidth]{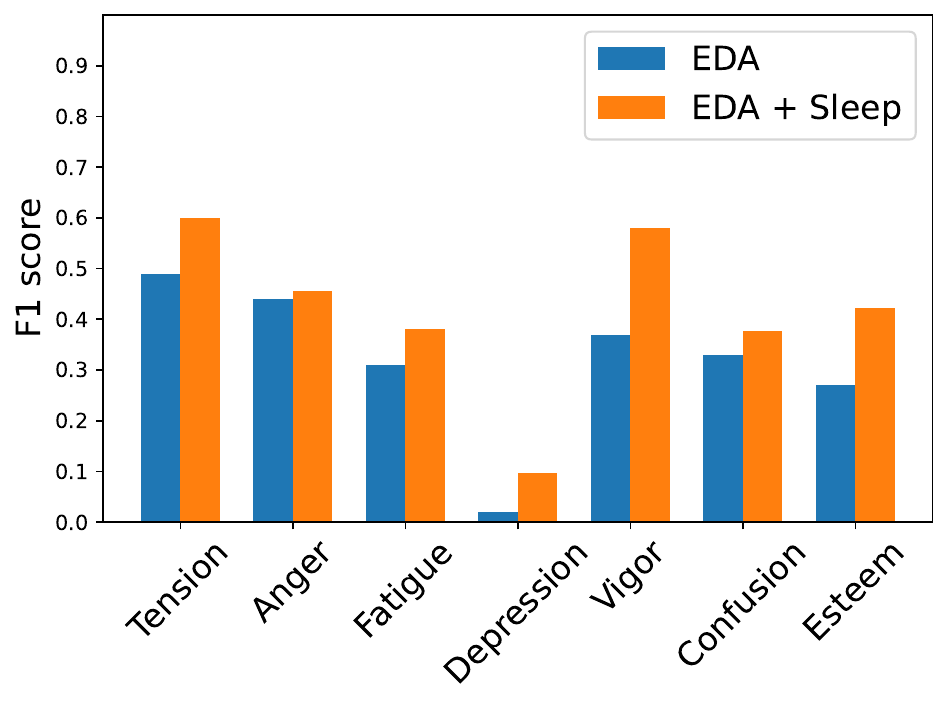} &
        \includegraphics[width=0.3\linewidth]{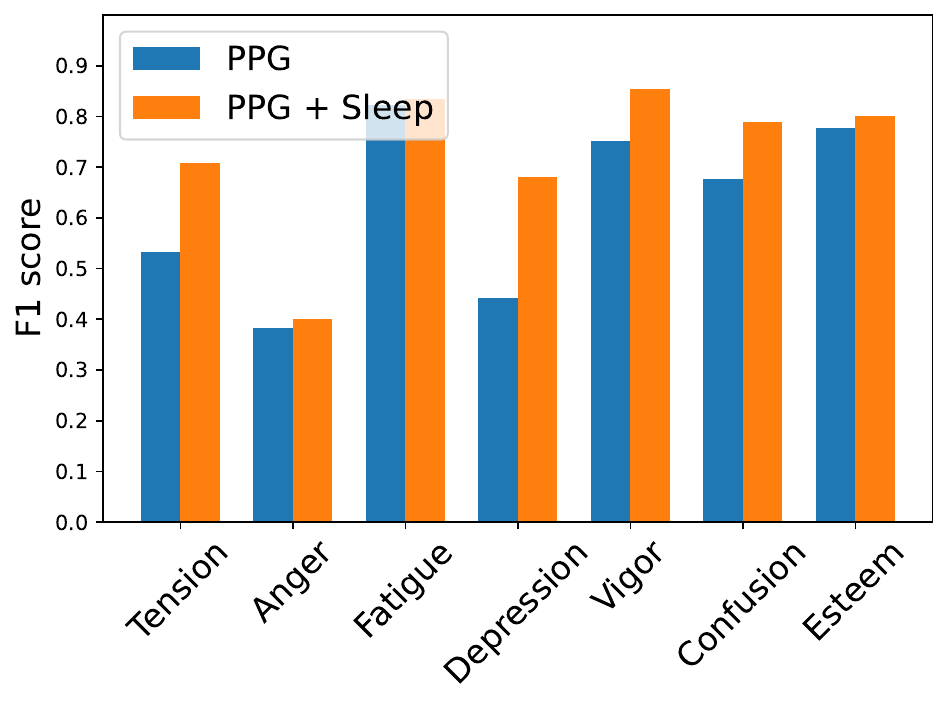} &
        \includegraphics[width=0.3\linewidth]{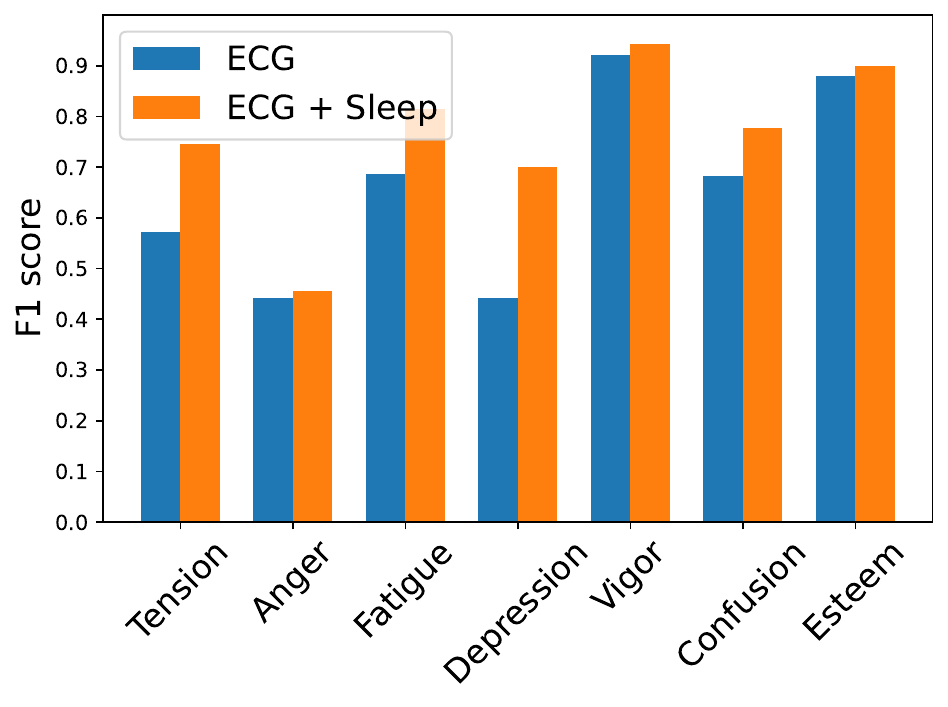} \\
    \end{tabular}
    \caption{Impact of sleep on classification of different classes of mood with EDA (left), PPG (middle), and ECG (right).}

    \label{fig:mood-level}
\end{figure*}

Next, we examine the effect of using sleep-related measures on the performance of our method. To this end, we use the frozen wearable time-series encoder and perform linear evaluation without the rest of our method which uses the sleep-related measures. We present the results in Table \ref{table:modalities}, which shows that using sleep-related measures as a complimentary modality leads to improvements across all three wearable time-series. In ECG-based mood recognition, adding sleep-related measures provides a performance boost of 9\% in F1 and 7\% in accuracy. With PPG, we observe boosts of 9\% and 12\% in F1 and accuracy respectively. Lastly for EDA, we observe a similar trend where 11\% and 10\% improvements are obtained for in F1 and accuracy respectively.

\begin{table}[t]
\centering
\small
\caption{Performance of our proposed method.}
\begin{tabular}{llccc}
\toprule
\textbf{Modality} & \textbf{Method} & \textbf{F1} & \textbf{Acc.}\\\midrule\midrule
\multirow{5}{*}{EDA + Sleep} & \citeauthor{hong2020holmes} \cite{hong2020holmes}& 0.33 & 0.30 \\
& \citeauthor{zihlmann2017convolutional} \cite{zihlmann2017convolutional}  & 0.29 & 0.27\\
& \citeauthor{chen2021signal} \cite{chen2021signal} & 0.31 & 0.30\\
& \citeauthor{behinaein2021transformer} \cite{behinaein2021transformer}  & 0.40 & 0.36\\
 &\textbf{Ours} & \textbf{0.42} & \textbf{0.39}\\
\hline
\multirow{5}{*}{PPG + Sleep} &  \citeauthor{hong2020holmes} \cite{hong2020holmes} & 0.48 & 0.46 \\
& \citeauthor{zihlmann2017convolutional} \cite{zihlmann2017convolutional}  & 0.44 & 0.42\\
& \citeauthor{chen2021signal} \cite{chen2021signal} & 0.52 & 0.50\\
& \citeauthor{behinaein2021transformer} \cite{behinaein2021transformer}  & 0.60 & 0.59\\
 &\textbf{Ours} & \textbf{0.72} & \textbf{0.69}\\
\midrule
\multirow{5}{*}{ECG + Sleep}& \citeauthor{hong2020holmes} \cite{hong2020holmes} & 0.56 & 0.52 \\
& \citeauthor{zihlmann2017convolutional} \cite{zihlmann2017convolutional}  & 0.49 & 0.47\\
& \citeauthor{chen2021signal} \cite{chen2021signal} & 0.50 & 0.46\\
& \citeauthor{behinaein2021transformer} \cite{behinaein2021transformer}  & 0.65 & 0.58\\
 &\textbf{Ours}   & \textbf{0.76} & \textbf{0.71}\\
\bottomrule
\end{tabular}
\label{table:multimodal}
\end{table}

\begin{table}[t]
\centering
\setlength \tabcolsep{20pt}
\small
\caption{Comparison of mood classification using wearable time-series with and without sleep-related measures.}
\begin{tabular}{lcc}
\toprule
\textbf{Modality} & \textbf{F1} & \textbf{Acc.}\\\midrule\midrule
EDA & 0.31 & 0.29 \\
EDA + Sleep & 0.42 & 0.39 \\
\midrule
PPG & 0.63 & 0.57\\
PPG + Sleep & 0.72 & 0.69 \\
\midrule
ECG & 0.67 & 0.64\\
ECG + Sleep  & 0.76 & 0.71 \\
\bottomrule
\end{tabular}
\label{table:modalities}
\end{table}

We then investigate the performance breakdown for different classes of mood, with and without the sleep-related measures, and present the results in Figure \ref{fig:mood-level}. 
Overall, we observe that the addition of sleep boosts performance for the classification of every mood class regardless of the type of wearable time-series used.
Next, it can be seen that using EDA (without sleep), tension and anger are classified with the highest F1 scores, whereas the addition of sleep results in the highest boosts for vigor, followed by esteem.
On the other hand, when using PPG (without sleep), the highest F1 scores are achieved for fatigue and esteem recognition, while the addition of sleep results in the highest boosts for depression and vigor.
Lastly, using ECG (without sleep) leads to the highest F1 scores for vigor and esteem, whereas the highest improvements by adding sleep are achieved for depression and tension. 
These findings are aligned with our understanding of the physiological connections between sleep and depression, as documented in prior research \cite{castiglione2023sleep, zhang2023psychosocial}. These studies have demonstrated that the quality of sleep from the previous night can significantly impact the onset of depression.

\begin{figure}[t!]
    \centering
    \includegraphics[width=1\columnwidth]{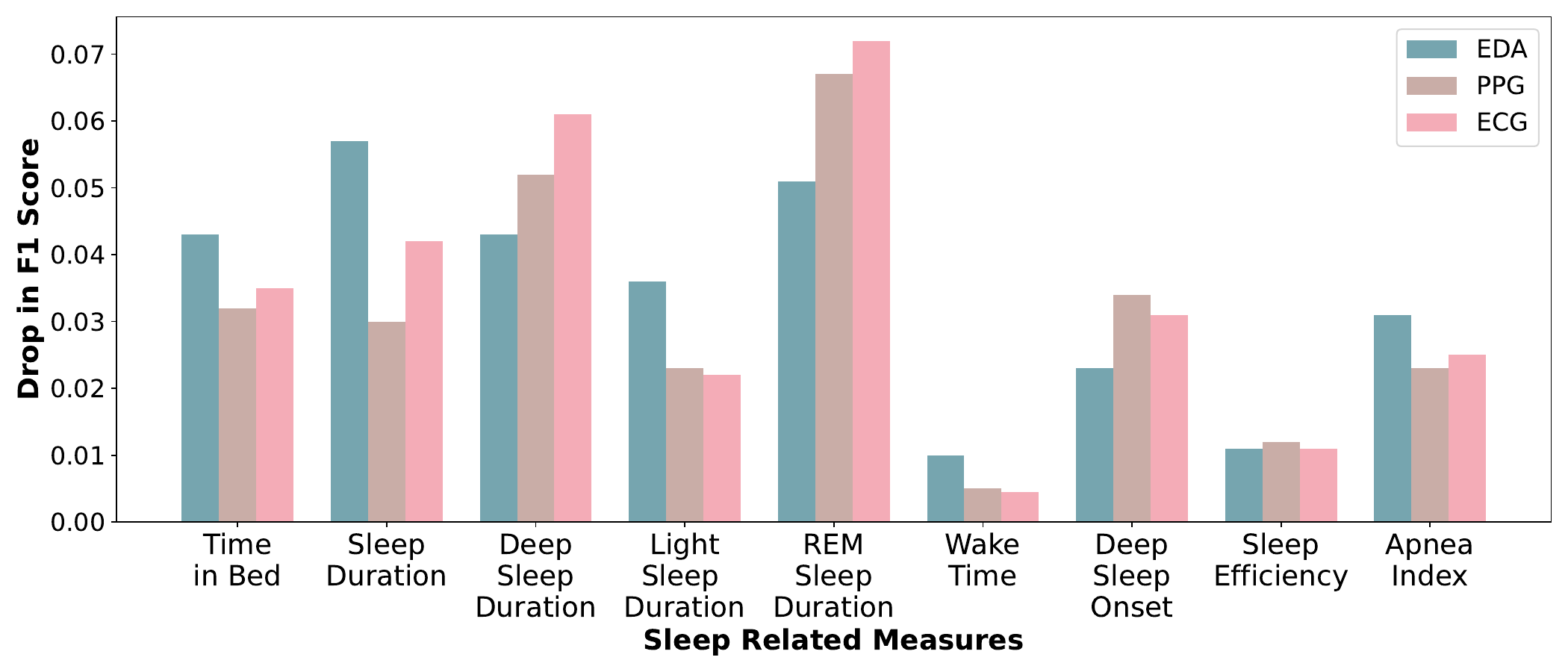}
    \caption{Ablation of individual sleep-related measures.}
    \label{fig:sleep-ablation}
\end{figure}

\begin{figure}[t]
    \centering
\includegraphics[width=0.83\linewidth]{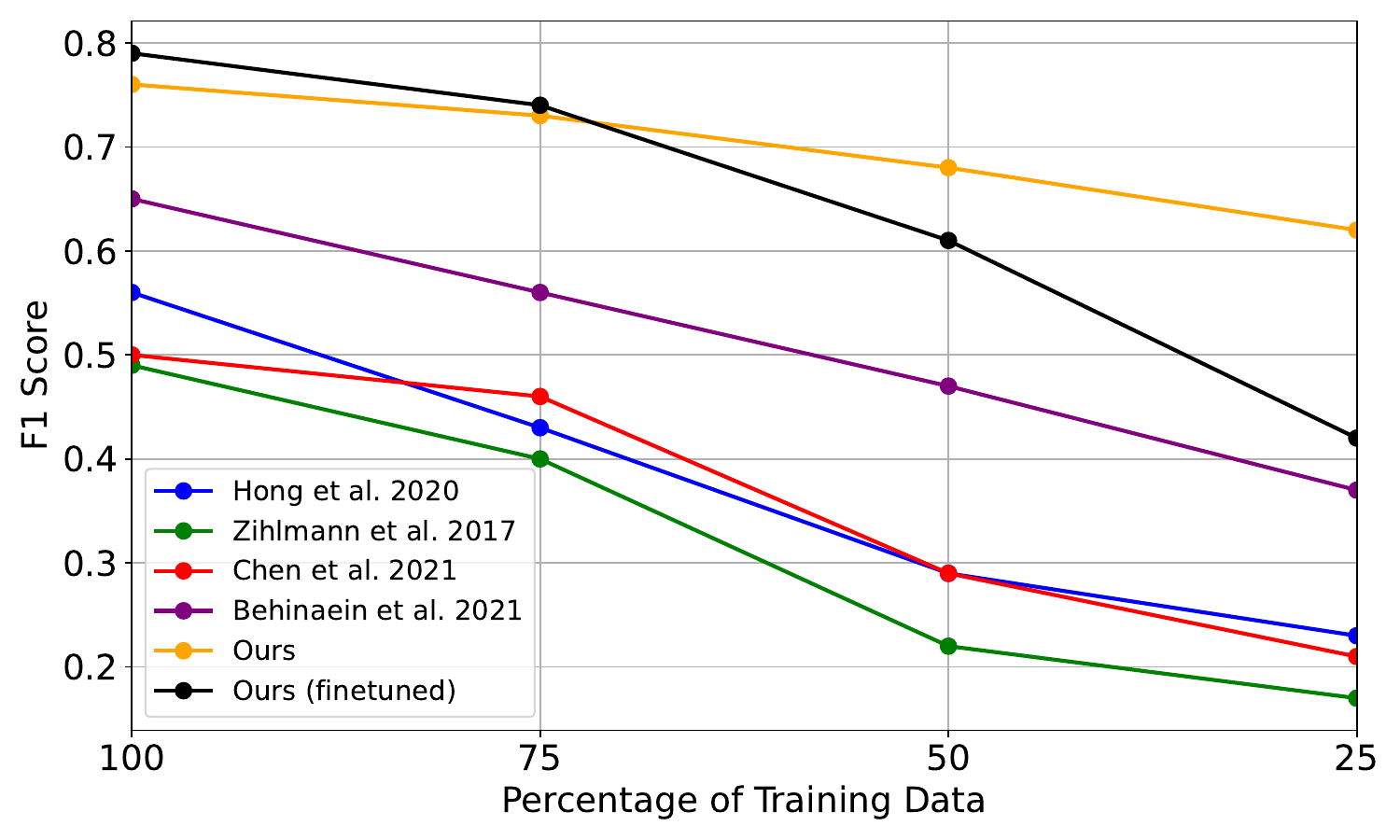}
    \caption{Impact of the amount of data used for training with ECG+Sleep.}
    \label{fig:sample-efficiency}
\end{figure}

Furthermore, we analyze the impact of different training methods on mood recognition performance in Table \ref{table:ablation}, specifically full-finetuning, pre-training, and NapTune. For full-finetuning, we update all the weights of the pre-trained encoder with sleep-related measures paired with wearable signals as input, without adding any prompt parameters. We observe an improvement of 3\% to 4\% in the case of finetuning the ECG and PPG-based models across both F1 and accuracy, while an increase of 8\% and 9\% respectively in the case of the EDA-based model. Our intuition suggests that the pre-training of EDA is not as strong as PPG and ECG counterparts, particularly due to a lesser amount of pre-training data. This in turn results in such an improvement in performance with finetuning compared to NapTune which uses Linear protocol. Upon training the model from scratch without any pre-trained weights, we observe a similar trend for all the modalities, with a performance drop ranging between 2\% to 4\% in both metrics.

Next, we study the contribution of individual sleep-related measures on mood recognition performance. Specifically, we ablate each sleep-related measure through masking at test time, and evaluate the model's performance. As shown in Figure \ref{fig:sleep-ablation}, REM sleep duration has the highest impact on mood recognition, followed by deep sleep and sleep duration. These findings are well-aligned with our understanding of how REM and deep sleep influence emotion regulation. Several studies \cite{genzel2015role, hutchison2015role, wiesner2015effect, walker2009overnight, goldstein2014role} have highlighted that during REM sleep the brain strengthens important emotional memories, which influences emotional regulation after waking up. Moreover, deep sleep has been highlighted as primarily responsible for consolidating and stabilizing emotional memories \cite{cairney2015complementary, cairney2014targeted}. Accordingly, both factors are highly influential in regulation of emotions and mood, which conforms with our observations from this experiment.

Lastly, we analyze the trend of mood recognition performance based on the amount of training data used for all the methods. As shown in Figure \ref{fig:sample-efficiency}, our proposed NapTune shows a much less steep decrease in performance when reducing training data, with up to a 62\% F1 score when we use only 25\% of the training data. In contrast, the other baselines show a much steeper descent, especially when reducing the training data from 50\% to 25\%. 

\begin{table}[t]
\centering
\small
\caption{Performance of different training methods.}
\begin{tabular}{clcc}
\toprule
\textbf{Modality} & \textbf{Variants} & \textbf{F1} & \textbf{Acc.}\\\midrule \midrule
\multirow{3}{*}{EDA + Sleep} & Ours & 0.42 & 0.39\\
& w/ finetuning & 0.51 $(\uparrow 9\%)$ & 0.47 $(\uparrow 8\%)$\\
& w/o pre-training & 0.40 $(\downarrow 2\%)$ & 0.36 $(\downarrow 3\%)$\\\midrule
\multirow{3}{*}{PPG + Sleep} & Ours & 0.72 & 0.69\\
& w/ finetuning & 0.75 $(\uparrow 3\%)$ & 0.73 $(\uparrow 4\%)$\\
& w/o pre-training & 0.69 $(\downarrow 3\%)$ & 0.67 $(\downarrow 2\%)$\\
\midrule
\multirow{3}{*}{ECG + Sleep} & Ours & 0.76 & 0.71\\
& w/ finetuning &  0.79 $(\uparrow4\%)$ & 0.74 $(\uparrow3\%)$\\
& w/o pre-training & 0.72 $(\downarrow 4\%)$ & 0.68 $(\downarrow 3\%)$\\
\bottomrule
\end{tabular}
\label{table:ablation}
\end{table}

\section{Conclusion}

We introduced NapTune, a prompt-tuning framework that involves adapting a frozen Transformer encoder, by adding lightweight prompt parameters into each transformer layer, to efficiently utilize the previous night's sleep-related measures as complementary information alongside wearable time-series data. Our results demonstrate that incorporating sleep data significantly improves mood classification, achieving up to an 11\% increase in F1 scores over baseline methods that only use wearable data. Our proposed NapTune framework outperforms several state-of-the-art multimodal baselines by a minimum margin of 11\%, 12\%, and 2\% for ECG, PPG, and EDA respectively. Future research directions may include studying other approaches to using pre-trained encoders, such as adaptors. Additionally, the notion of using generative models to fill the gap in instances where previous night's sleep measures may be missing, can be explored.

\printbibliography
\end{document}